\documentstyle[prb,aps,epsf]{revtex}

\tighten

\begin{document}

\twocolumn[\hsize\textwidth\columnwidth\hsize\csname @twocolumnfalse\endcsname

\title{Structural, electronic, and dynamical properties of amorphous gallium
arsenide: a comparison between two topological models}

\author{Normand Mousseau\cite{mousadd} and Laurent J. Lewis \cite{lewadd}}

\address{D{\'e}partement de physique and Groupe de recherche en physique
et technologie des couches minces (GCM),
Universit{\'e} de Montr{\'e}al, Montr{\'e}al, Qu{\'e}bec, Canada, H3C 3J7}

\maketitle

\begin{center}
Submitted to the Physical Review B - February 1997 \\
\end{center}

\begin{abstract}
We present a detailed study of the effect of local chemical ordering on the
structural, electronic, and dynamical properties of amorphous gallium
arsenide. Using the recently-proposed ``activation-relaxation technique'' and
empirical potentials, we have constructed two 216-atom tetrahedral continuous
random networks with different topological properties, which were further
relaxed using tight-binding molecular dynamics. The first network corresponds
to the traditional, amorphous, Polk-type, network, randomly decorated with Ga
and As atoms. The second is an amorphous structure with a minimum of wrong
(homopolar) bonds, and therefore a minimum of odd-membered atomic rings, and
thus corresponds to the Connell-Temkin model. By comparing the structural,
electronic, and dynamical properties of these two models, we show that the
Connell-Temkin network is energetically favored over Polk, but that most
properties are little affected by the differences in topology. We conclude
that most indirect experimental evidence for the presence (or absence) of
wrong bonds is much weaker than previously believed and that only direct
structural measurements, i.e., of such quantities as partial radial
distribution functions, can provide quantitative information on these defects
in {\it a}-GaAs.
\end{abstract}
\pacs{PACS: 61.43.Dq,61.43.Bn,63.50.+x,71.55.Jv}

\vskip2pc]
\narrowtext

\section{Introduction}

After twenty five years of effort, the structure of amorphous materials, and
how it affects the electronic and vibrational properties, remains largely
unresolved. Most experimental probes yield information that is averaged out
over rather large lengthscales and therefore lack the sensitivity to
discriminate between various possible structural models of the same material.
Techniques such as EXAFS, while they can provide structural information at
the atomic level, are often too imprecise to yield definite and unambiguous
structural parameters. One must therefore proceed iteratively between models
and experimental data in order to acquire the desired structural information.

In the case of amorphous semiconductors, progress in the development of
satisfactory structural models has been hindered by difficulties in
constructing ``continuous random networks'' (CRN) with different topologies,
i.e., appropriate to different materials. The idea of representing the
structure of amorphous semiconductors by CRN's was first proposed by
Zachariasen;\cite{zachariasen32} in this picture, the material is assumed to
consist of a ``collage'' of tetrahedra quite similar to those found in the
corresponding crystal but randomly connected through their vertices. Based on
these ideas, the mechanical CRN constructed by Polk was found to provide a
very satisfactory description of the topology of elemental amorphous
semiconductors.\cite{polk71}

For compound materials, however, the situation is not as clear: The
building-block tetrahedra, if they exist, can be formed in many different
ways, depending on the chemical identities of the atoms, the arrangement of
which is determined by the bonding characteristics of the material. For
example, in the case of the III-V compound GaAs, chemical ordering should
predominate because the material is partly ionic, i.e., the number of bonds
between like atoms (``wrong bonds'') should be minimal. Ideally, a
tetrahedron should consist of a Ga atom surrounded by exactly four As atoms
(or vice versa), as is the case in a perfect (zinc-blende) GaAs crystal. The
actual structure of the disordered material, therefore, will be determined by
a balance between the cost of elastically deforming the network while
maintaining perfect chemical ordering and the cost of introducing wrong
bonds. In view of this, {\it a}-GaAs appears to be an ideal candidate for the
realization of the Connell-Temkin model\cite{connell74} --- a CRN similar to
Polk's but without odd-membered atomic rings. (A ring is defined as a closed
path between an atom and itself through a series of bonds). Indeed, in an
unconstrained CRN, there are inevitably both even- and odd-membered rings.
Since odd-membered rings necessarily bring about wrong bonds, which cost
Coulomb energy, it is expected that the number of them will be minimal in
{\it a}-GaAs, and ideally none as in the Connell-Temkin model if the cost in
elastic deformation energy associated with this constraint is not too large.

As noted above, detailed experimental information regarding the structure of
compound materials is in general difficult to obtain; this is particularly so
in the case of {\it a}-GaAs because of the close similarity between the
constituent elements. (They are near neighbors in the Periodic Table; in
fact, the similarity in size is another reason why the topology of {\it
a}-GaAs should correspond closely to that of the single-component
Connell-Temkin model.) Experimental evidence for the presence (or absence) of
wrong bonds in this material is indeed essentially non-existent, while the
question does not arise in elemental amorphous semiconductors, such as {\em
a}-Si. It is therefore important, in order to understand on a local scale the
topology of amorphous semiconductors, to examine idealized representations of
the two different types of networks, viz.\ Connell-Temkin and Polk, and
compare them with experiment. This is the object of the present paper.

In a recent article, we have shown that it is quite possible to construct a
binary-compound network which contains essentially no odd-membered rings
(i.e., wrong bonds), corresponding to an infinite Connell-Temkin
model;\cite{mousseau97} for GaAs, this model is found to be energetically
preferred over the Polk model, where both even- and odd-membered rings are
present. Likewise, for elemental and/or non-ionic tetrahedral semiconductors,
the Polk-type model is preferred for entropic reasons: elastic costs
associated with the constraint on the absence of odd rings are negligible but
the limitation on rings restricts significantly the number of possible
configurations. In this way, we have been able to achieve a direct structural
comparison of materials differing on the intermediate lengthscale --- {\em
a}-Si and {\em a}-GaAs.

In the present paper, we extend this study and examine in detail the
structural, vibrational, and electronic properties of amorphous GaAs as
described by the two types of CRN's mentioned above. This provides unique and
much-needed information on the effects of topology on the properties of
amorphous semiconductors. Indeed, upon comparing different networks
constructed using the same energy scheme, it is possible to isolate, in
measured properties, those effects arising from topology from those due to
the interactions between atoms --- evidently something which cannot be done
experimentally.

\section{Methodology}

The timescale on which chemical ordering takes place when a binary compound
is cooled down from the liquid phase depends strongly on the ionicity of the
material. For example, molecular-dynamics (MD) simulations on a timescale of
picoseconds are sufficient to ensure proper ordering of silica
(SiO$_2$),\cite{sarnthein95} which has a Phillips ionicity of
2.09.\cite{phillips73} In contrast, corresponding simulations of {\em
a}-GaAs, which is only slightly ionic (0.22), have not given clear indication
of chemical ordering. Whether these results reflect the actual structure of
GaAs or limitations of MD simulations needs to be addressed using a different
approach for constructing structural models.

One possibility is to bypass the dynamics of formation and devise an
appropriate static structure optimization scheme; by ``appropriate'' we mean
which will lead to a physically realistic structural model. The route
followed need not be physical, however: the philosophy we adopt here is one
where ``the end justifies the means''. This argument will become evident in
the discussion which follows, but one closely-related precedent can be
invoked --- the celebrated Wooten-Winer-Weaire (WWW) algorithm for
constructing models of {\em a-}Si:\cite{wooten85} In this approach,
crystalline Si is amorphized through a sequence of bond-switching moves which
are totally unphysical; yet, the final (converged) structure possesses much
of the properties of real {\em a}-Si. The method cannot be employed in the
case of compound semiconductors because it is the essence of it to introduce
odd-membered rings and wrong bonds. (It involves breaking nearest-neighbor
bonds and forming second-neighbor --- thus wrong --- bonds.) These
limitations of the WWW algorithm, and/or absence of alternative models, have
significantly hindered the study of amorphous semiconductors over the last
ten or so years.

The activation-relaxation technique (ART) recently proposed by Barkema and
Mousseau\cite{barkema96} provides a way to circumvent the restrictions of the
WWW model:\cite{wooten87} Given a model of interatomic potentials, which can
be of any form --- from purely empirical to fully {\it ab initio}, the method
(which is event-based) forces relaxation through a series of {\em physical}
moves in a configurational space reduced to a set of isolated energy minima
connected together by paths going through saddle points on the
configurational energy landscape. Since the moves are defined in the
$3N$-dimensional configurational space, where $N$ is the number of atoms, the
algorithm is completely independent of the structure in the 3-dimensional
real space. A move, therefore, can involve any number of atoms, and in
particular be local or span the whole system; it is not limited to a
pre-determined list of events, such as bond switches {\em {\`a} la} WWW or atomic
exchanges. Further, since the configurational space is reduced to an
(infinite) number of discrete points, events can be defined uniquely.

Full details and illustrations of the ART method can be found in Ref.\
\onlinecite{barkema96}; here we give a brief overview. An ART simulation
starts with the system in a local minimum of the potential-energy surface.
The configuration is then placed slightly out of equilibrium by operating a
small change on the system, e.g., moving an atom in a random direction by a
very small amount. The component of the force {\em parallel} to the
displacement of the configuration is then inverted and the whole
configuration pushed away from the nearby minimum following the {\em modified
force}:
   \begin{equation}
   \vec{G}=\vec{F}-(1+\alpha) (\vec{F}\cdot \Delta \hat{X}) \Delta \hat{X},
   \label{eq:art}
   \end{equation}
where $\vec{G}$ and $\vec{F}$ are both $3N$-dimensional vectors, $\alpha$ is
a dimensionless parameter (here set to 0.15) and $\hat{X}$ is the unit-vector
displacement from the local minimum to the current position. Equation
\ref{eq:art} is iterated until the modified force (and thus also the real
force) vanishes, indicating that a saddle point has been reached. The
configuration is then pushed slightly away from the saddle point and relaxed
to a new local minimum. This procedure is iterated until convergence, i.e.,
until no further changes in the total energy are observed.

A few remarks are in order: (i) Since the algorithm is defined in
configurational space, there is no limit to the number of atoms which can be
involved in a single event; for the materials studied here, events involving
one to a few tens of atoms have been observed. (ii) In order to prevent the
configuration from moving to close-by saddle points which are not significant
at finite temperature, a repulsive force is introduced that excludes a region
of width $\sim x_c$ around any local minimum:
   \begin{equation}
   F_{rep} = A (x-x_c);
   \end{equation}
here $x$ is the scalar displacement of the configuration from the local
minimum, $x_c$ is a cutoff parameter, and $A$ is the strength of the
repulsive part. Both $x_c$ and $A$ are drawn, for each new event, from a
linear random distribution: $0.3 < x_c < 1.3$ \AA\ and $0.5 < A < 1.5 $
eV/\AA$^2$. (iii) After each event, the volume of the system is optimized
such as to minimize the configurational energy. (iv) Both activation to the
saddle point and relaxation to the nearest local minimum are performed using
the Levenberg-Marquardt algorithm;\cite{nr} this algorithm, which includes
both the steepest-descent and the Hessian approaches, is fairly efficient
around a minimum and does not misbehave far from it. (v) The optimization is
performed using a multiple-configuration simulated-annealing approach; thus,
each new event is accepted with a probability
   \begin{equation}
      P_{accept}(\{x_{i+1}\}) = \exp \left[
         -\frac{E(\{x_{i+1}\})-E(\{x_i\})}{k_B T} \right],
   \end{equation}
where here the temperature is a is non-physical parameter. In practice, two
configurations are run in parallel at different temperatures. After each
step, the energies are compared and the configurations are switched according
to a Metropolis rule. The use of multiple configurations allows for more
efficient sampling of configurational space, permitting configurations to
escape from dead-end minima, i.e., those which cannot lead directly to a
lower-energy state.

The force $\vec{F}$ in Eq.\ \ref{eq:art} is derived from an interatomic
potential which, as noted above, can in principle be of any form. Evidently,
the final, optimized structure will depend on the choice of potential. With
respect to GaAs, there exists (to our knowledge) no satisfactory empirical
potential for the disordered phase. Our own attempts in this regard using a
recently-proposed Tersoff-type potential\cite{tersoff89} have lead to
structures in deep disagreement with experiments. For the present work,
construction of the computer models proceeded in two stages under two sets of
potentials: first, ART optimizations using modified Stillinger-Weber
potentials (see below) were carried out; and second, the models were further
relaxed under semi-empirical tight-binding (TB) potentials. The reason for
this ``double-relaxation'' approach is that while ART relaxations can in
principle be done with the TB potentials, this remains a complicated and
computer-intensive enterprise. Carrying out a ``first pass'' with empirical
potentials allows optimized structures to be obtained rapidly, while the
final TB relaxation provides a physically-meaningful basis for the models;
indeed, we have found the properties of our networks to be only little
affected by the final TB relaxation, thus indicating the convergence and
validity of the procedure.

\section{Model preparation}

As discussed in the Introduction, we consider here two networks with
different topologies; following Ref.\ \onlinecite{mousseau97}, these are
poetically labeled CRN-A and CRN-B. CRN-A corresponds to a Polk-type network,
while CRN-B is Connell-Temkin like. The initial ART relaxation was performed,
in both cases, with a Stillinger-Weber potential for
silicon,\cite{stillinger85} except that, in order to compensate for too weak
an angular force in the original set of parameters,\cite{barkema96} the
three-body contribution was increased by 50\%. For CRN-B, moreover, a
repulsive term between like atoms was introduced in order to minimize the
number of wrong bonds and favor a Connell-Temkin like topology:
   \begin{equation}
   E_{rep}=\sum_{<ij>}\epsilon A_{ij} \left[ 1 + \cos\left(\pi
   \frac{r_{ij}}{s}\right)\right],
   \end{equation}
where $\epsilon$ is the energy parameter of the Stillinger-Weber potential,
$A_{ij}=1.2$ for like atoms and zero otherwise, and $s=3.6$ \AA; this is also
the value for the cut-off of this potential. Except for this additional
repulsive energy, both atomic species are treated in exactly the same way for
CRN-B. The validity of these modifications is assessed {\it a posteriori}
when the lattice is further relaxed with a physical TB potential (see below).
As with any static approach to modeling complex systems, it is the final
structure which determines the quality of the method, i.e., here, the end
justifies the means, as noted earlier.

The ART optimization was initiated, in each case, from a 216-atom
random-packed configuration at the {\it c}-Si density; periodic boundary
conditions were employed in order to eliminate surface effects. Iteration of
the ART procedure was carried out until changes in the total energy were
judged insignificant. For CRN-A, the optimization is fairly rapid and takes
two or three days on a R8000 computer. In the case of CRN-B, considerable
atomic diffusion is required in order to minimize the number of wrong bonds,
and the simulation takes longer --- about one week. These run times should be
contrasted with the several months needed for a TB simulation of a 64-atom
system on a workstation or a Car-Parrinello\cite{car85} (CP) simulation on a
state-of-the-art parallel supercomputer.

{\em Both} CRN-A and CRN-B were then relaxed with {\em both} the
Goodwin-Skinner-Pettifor (GSP) TB potential for Si\cite{goodwin89} {\it and}
the Molteni-Colombo-Miglio (MCM) TB potential for
GaAs.\cite{molteni94a} Thus, in total, we have four different
zero-temperature TB-relaxed models: CRN-A-Si, CRN-A-GaAs, CRN-B-Si, and
CRN-B-GaAs, i.e., for each material, two models with different topologies. In
the case of CRN-A-GaAs, it is necessary to ``label'' the atoms after the ART
simulation and before the TB relaxation. This was done using a random
``label-switching'' procedure designed to minimize the number of wrong bonds
on the lattice. In this way, we obtained a proportion of wrong bonds of 14\%,
as close as one can possibly get, for a finite-size system, to the
theoretical value of 12\% for optimal ordering on a Polk-type
CRN.\cite{theye80} The configurational energies of each of the four models
after the static TB relaxation are presented in Table \ref{tab:energies}. A
detailed comparison between Si and GaAs is given elsewhere\cite{mousseau97}
and we therefore only discuss, here, the two models for GaAs, viz.\
CRN-A-GaAs and CRN-B-GaAs, hereafter referred to simply as CRN-A and CRN-B.

Following the static TB simulation, the two samples were further relaxed at
300 K using MD for a total of 7.0 ps. Although this is a fairly short period
of time, it is enough to ensure that both CRN-A and CRN-B have evolved into
deep local minima of the potential energy surface. In order to verify this,
we have also annealed the samples at 700 K during 8.8 ps before running again
at 300 K for 3.5 ps and 10 K for 0.9 ps. The CRN-B network was found to be
only very weakly affected by the high-temperature annealing. For CRN-A, in
contrast, annealing resulted in significant changes in the topology; the
average coordination, for instance, increased from 3.95 to 4.19. Likewise,
the energy {\em increased} from $-$13.45 eV/atom to $-$13.39 eV/atom. This in
fact could be expected: it indicates that model CRN-A is not a proper state
for GaAs. Though it would take much longer runs to find out, it is very
likely that the system is trying to find a route (through a higher-energy
transition state) towards the preferred configuration --- one without wrong
bonds. Thus, these high-temperature simulations confirm that CRN-B is indeed
a much better model for the structure of {\it a}-GaAs than CRN-A. In the
discussion that follows, the structures relaxed at 300 K before annealing
will be used for comparing the two models.

As noted above, here we used 216-atom unit cells, the largest size we could
deal with \hfill  in \hfill our \hfill  TB \hfill simulations,

\begin{table}
\caption{
Energies in eV per atom of the two networks statically relaxed using the two
sets of tight-binding parameters (see text). The atoms on the CRN-A model
have been placed in such a way as to minimize the number of wrong bonds. Also
given are the energies of the crystalline phases. SL refers to the
tight-binding simulation of Seong and Lewis, Ref.\
\protect\onlinecite{seong96}
}
\begin{tabular}{l|ccc}
Network & \multicolumn{2}{c}{TB parameters} & \\
        &   Si      &  GaAs     &  GaAs (SL) \\ \hline
  CRN-A & $-$13.172 & $-$13.450 &            \\
  CRN-B & $-$13.163 & $-$13.561 &  $-$13.450 \\
Crystal & $-$13.389 & $-$13.802 &  $-$13.802 \\
\end{tabular}
\label{tab:energies}
\end{table}

\vspace*{-2.5cm}
\begin{figure}
\epsfxsize=14cm
\epsfbox{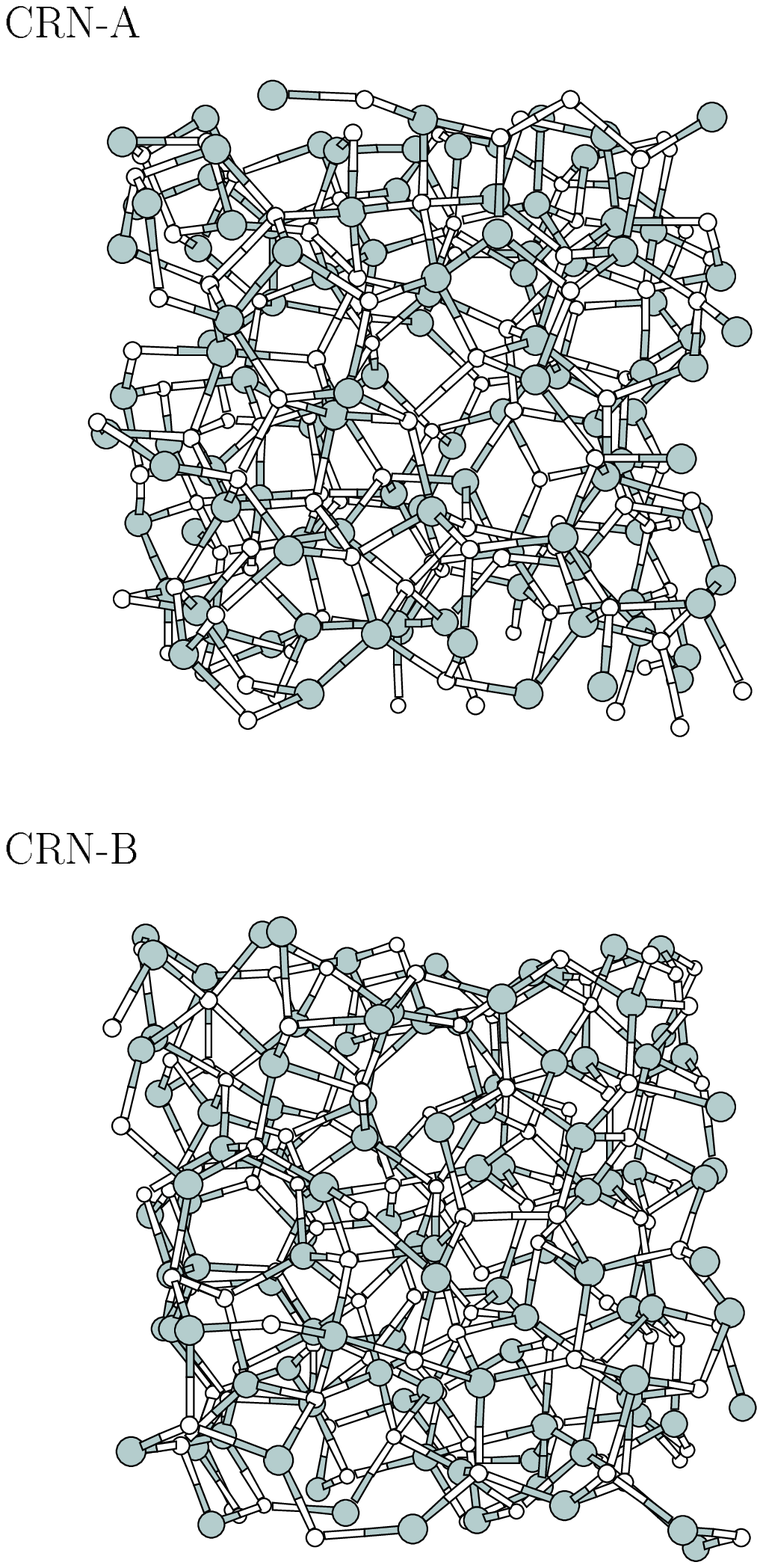}
\vspace*{-1.cm}
\caption{
Ball-and-stick representation of CRN-A and CRN-B after relaxation with a GaAs
TB potential; small white circles are As and larger grey circles are
Ga atoms.}
\label{fig:mol}
\end{figure}

\noindent nevertheless large enough to provide a
satisfactory description of amorphous semiconductors. With 64-atom cells, we
found ART to lead, in all cases, to the crystalline state, indicating that
the configurational space for a system of this size is small enough for ART
to find the global minimum. In contrast, for the 216-atom systems,
crystallization never occured and so we conclude that the local minimum we
find are truly optimized.

\section{Structural properties}

In Fig.\ \ref{fig:mol} are shown ball-and-stick representations of the two
{\it a}-GaAs samples. For CRN-B, wrong bonds are few but one can be seen in
the top right quadrant where its presence gives rise to both a five- and a
seven-membered ring. It is clear from this figure that the two models, though
clearly ordered at short-range --- both topologically and chemically --- bear
no trace of crystallinity. (It is often the case that partly crystalline
samples cannot be recognized, because of averaging and thermal agitation, in
such quantities as radial distribution functions and static structure
factors.) In fact, based on this visual inspection, it is already evident to
a trained eye that the samples are excellent models of the amorphous
material.

We now proceed with a more quantitative analysis of the models. First we
give, in Table \ref{tab:coord}, some of the usual system-averaged
structural quantities --- coordination numbers, proportion of wrong bonds,
and width of the bond-angle distribution --- for our two models, both at 0
and at 300 K.  Here we do not distinguish between the various types of
correlations and treat all atoms on the same footing. The reason for this
is that the identity of the atoms prior to the ART relaxation was
introduced in an ad-hoc fashion; since the TB relaxation does not change
the connectivity of the network, the two species are still topologically
equivalent in the final configuration. This, it turns out, is consistent
with previous simulations by Molteni {\it et al.}\cite{molteni94b} and by
Fois {\it et al.}\cite{fois92} which show the two species to behave
symmetrically, and is also supported by experiment, which shows As and Ga
to be both four-fold coordinated (taking due account of variations in
composition).\cite{udron92}

We first observe that the structural characteristics of both CRN-A and CRN-B
satisfy the requirements of a ``good'' CRN, namely four-fold coordination and
small bond-angle deviation. From Table \ref{tab:coord}, it is apparent that
CRN-A and CRN-B have a density of coordination defects lower than that of
previous models of {\it a}-GaAs. In particular, both models have a
coordination of almost exactly four, i.e., most atoms are perfectly
coordinated but a few are under-coordinated (cf.\ Table \ref{tab:coord}). [In
order 

\begin{table}[t]
\caption{
Structural characteristics of the models discussed in the text, at both 0 K
and 300 K: distribution of coordination numbers, $Z$ (and nearest-number
cutoff distance, $r_{NN}$), density of wrong bonds, and width of the
bond-angle distribution, $\Delta\theta$. Also given are the corresponding
numbers from other simulations, all at 0 K: SL --- tight-binding simulations
of Ref.\ \protect\onlinecite{seong96}; MCM --- tight-binding simulations of
Ref.\ \protect\onlinecite{molteni94b}; CP --- Car-Parrinello simulations of
Ref.\ \protect\onlinecite{fois92}.
}
\begin{tabular}{l|ccccccc}
                      & \multicolumn{2}{c}{CRN-A} & \multicolumn{2}{c}{CRN-B} & SL & MCM & CP \\
                      & 0 K   & 300 K & 0 K   & 300 K &       &      &       \\ \hline
$Z=$ 3                & 0.046 & 0.128 & 0.051 & 0.118 & 0.242 & 0.14 & 0.219 \\
$Z=$ 4                & 0.954 & 0.845 & 0.944 & 0.830 & 0.598 & 0.66 & 0.781 \\
$Z=$ 5                & 0     & 0.026 & 0.005 & 0.045 & 0.129 & 0.18 & 0     \\
$Z=$ 6                & 0     & 0.001 & 0     & 0.004 & 0.024 &      & 0     \\
$Z=$ 7                & 0     & 0.000 & 0     & 0.002 & 0.007 &      & 0     \\
$<Z>$                 & 3.95  & 3.90  & 3.95  & 3.95  & 3.94  & 4.09 & 3.83  \\
$r_{NN}$ (\AA)        & 3.0   & 3.1   & 3.0   & 3.1   & 3.0   & 3.0  & 2.8   \\
Wrong bonds (\%)      & 14.1  & 14.2  & 3.9   & 5.2   & 12.2  & 12.9 & 10.0  \\
$\Delta\theta$ (deg.) & 11.0  & 14.1  & 10.8  & 15.0  & 17.0  & 17.0 &       \\
\end{tabular}
\label{tab:coord}
\end{table}

\begin{figure}
\vspace*{-1.0cm}
\epsfxsize=8cm
\epsfbox{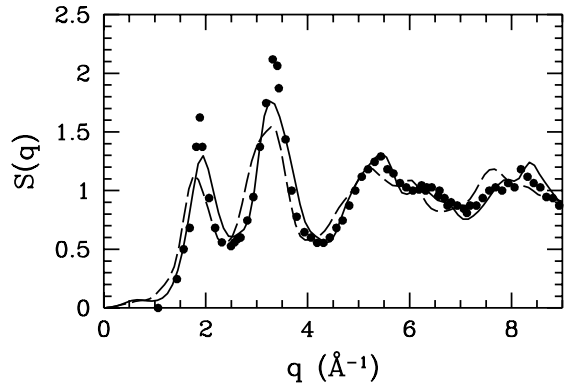}
\vspace*{-1.5cm}
\caption{
Total static structure factors for CRN-A (dashed line) and CRN-B (solid
line); the dots are the experimental data of Udron {\it et al.}, Ref.\
\protect\onlinecite{udron91}.
}
\label{fig:fss}
\end{figure}

\noindent
to define nearest neighbours, we use here the distance corresponding to
the minimum following the first maximum of the total radial distribution
function (see below), viz.\ 3.0 \AA.] Both networks, therefore, are
consistent with experiments.\cite{udron92} It is also clear from Table
\ref{tab:coord} that the finite-temperature models are totally equivalent to
the zero-temperture models: Thermal agitation only brings about a widening of
the distributions of neighbours and bond angles, leaving essentially
unchanged the number of wrong bonds.

The density of wrong bonds in model CRN-A is 14 or 15\%. As discussed
earlier, this value was obtained by assigning the identities of atoms on
CRN-A such as to minimize the number of wrong bonds, and corresponds quite
closely to the ``theoretical limit'' of 12\% for a Polk-type
CRN.\cite{theye80} It is also close to the values obtained in melt-and-quench
MD simulations of {\it a}-GaAs (10--13\% --- cf.\ Table \ref{tab:coord}). In
contrast, CRN-B, with less than 4 or 5\% of wrong bonds, is the closest
realization of an ``infinite'' Connell-Temkin model. Such a low proportion of
wrong bonds also seems to be in much better agreement with experiment, which
indicates that at most a few percent of wrong bonds are
present;\cite{senemaud85} as will be discussed below, however, the
measurements on which this estimate is based turn out to be much less
sensitive to the density of wrong bonds than what is usually believed.
Finally, visual inspection of Fig.\ \ref{fig:mol} reveals no spatial
concentration of wrong bonds; they appear to be homogeneously distributed on
both networks.

That CRN-B is a better model for the structure of {\em a}-GaAs than CRN-A is
evident from the configurational energies given in Table \ref{tab:energies};
it is also evident from this Table that the procedure employed here --- ART
plus TB-MD relaxation --- leads, in this case, to a better model than the
usual melt-and-quench MD approach. (Cf.\ the TB-MD results of Seong and
Lewis, Ref.\ \onlinecite{seong96}, given in Table \ref{tab:energies}.)

The ability of CRN-B to describe {\em a}-GaAs can also be assessed from the
total \hfill static \hfill structure \hfill factors (SSF's),

\begin{figure}
\vspace*{-0.5cm}
\hspace*{-1.4cm}
\epsfxsize=10cm
\epsfbox{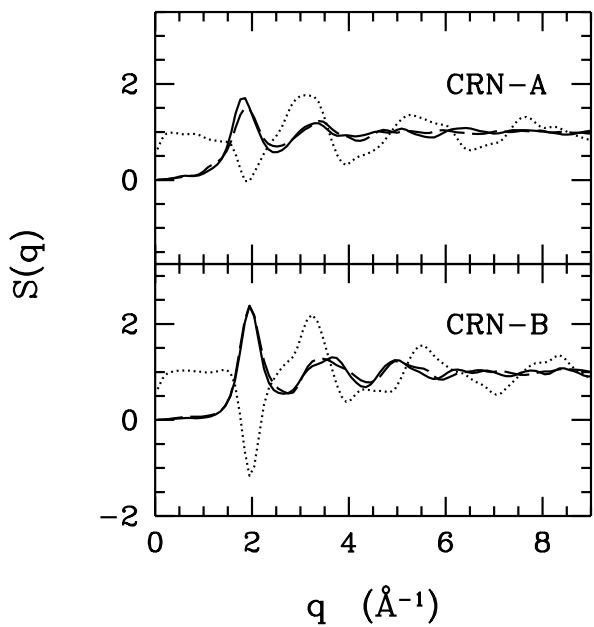}
\vspace*{-1.cm}
\caption{
Partial static structure factors for the two models, as indicated; the
dotted, dashed, and solid lines are for the Ga--Ga, Ga--As, and As--As
partial correlations.
}
\label{fig:fss2}
\end{figure}

\noindent 
presented in Fig.\ \ref{fig:fss}: the
SSF for CRN-B matches more closely the experimental data\cite{udron91} than
that for CRN-A. Unfortunately, the samples used in the experiment suffer from
some inhomogeneities; further structural measurements on better-quality
material would provide much-needed experimental data for more accurate
comparisons.

The partial SSF's for the two models are shown in Fig.\ \ref{fig:fss2}.
Although they differ in many ways, no evident signature of the presence of
wrong bonds in CRN-A can be identified: the differences between the two sets
of curves are essentially quantitative, and no peaks appear in one of the
partials which is totally absent in the other. However, as we discuss next,
the partial radial distribution functions (RDF's), obtained from the SSF's by
a Fourier transformation, allow a much better interpretation of these
differences.

\begin{figure}
\vspace*{-0.5cm}
\hspace*{-0.75cm}
\epsfxsize=10cm
\epsfbox{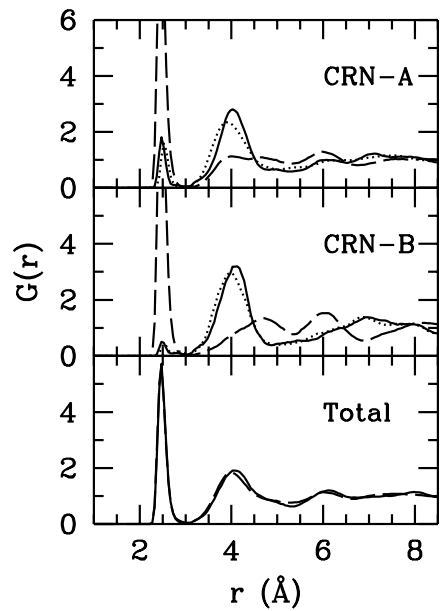}
\vspace*{-1.2cm}
\caption{
Partial radial distribution functions for the two models, as indicated; the
dotted, dashed, and solid lines are for the Ga--Ga, Ga--As, and As--As
partial correlations, respectively. The lower panel gives the total
(unweighted) radial distribution function for CRN-A (dashed line) and CRN-B
(solid line).
}
\label{fig:fdr}
\end{figure}

In effect, the partial RDF's can provide direct, quantitative, evidence of
the existence, and proportion, of wrong bonds; they are given in Fig.\
\ref{fig:fdr} for our models. The fact that CRN-B is chemically ordered is
clearly visible in the partial RDF's: The unlike-atom partial function,
Ga-As, exhibits a strong first-neighbour peak, but very little amplitude at
the second and fourth nearest-neighbour distances. In contrast, the
like-atom partial RDF's, Ga-Ga and As-As, have essentially no
nearest-neighbours, but exhibit strong second and fourth nearest-neighbour
peaks. Thus, chemical-order ``filters out'' the shell structure of the
material (on the short and intermediate lengthscales). As a result, the
large split peak of the Ga-As correlation function in the range 3.5-7 \AA,
which corresponds to third nearest-neigbbours, can be clearly isolated from
others. \hfill This is important because

\noindent this peak  
corresponds to the various
possible dihedral conformations; here we find that the two sub-peaks
correspond to dihedral angles of 60 and 180 degrees, as we indeed find
below through a direct calculation of the dihedral angle distribution.

These results points to the importance of measuring the partial RDF's (or
SSF's). Because Ga and As are close to one another, however, only the total
RDF is available from x-ray or other scattering measurements. This is true
also of EXAFS\cite{udron91} for which it is difficult to distinguish the
atomic type in the nearest-neighbor shell. This is unfortunate because, as
can be seen in Fig.\ \ref{fig:fdr}, most of the information on wrong bonds is
lost in the weighted sum over the partial RDF's: the total RDF's for the two
models are almost identical over the whole range of distances of interest, as
can be seen in Fig.\ \ref{fig:fdr}.

The quality of the models can also be inferred from the distributions of bond
and dihedral angles. In the case of {\em a}-Si, the width of the bond-angle
distribution at 0 K is found experimentally to lie in the range 10--12
degrees,\cite{connell74} in accord with recent, fully-optimized, WWW
models.\cite{djordjevic95} From MD simulations of {\em a}-Si, and more
recently {\em a}-GaAs, typical values for this quantity are in the range
15--17 degrees; in the case of {\em a}-GaAs, further, the distribution of
bond angles is observed to be asymmetric,\cite{molteni94b,seong96} biased
towards smaller angles, manifest of the presence of a significant number of
over-coordinated atoms. These results could be seen as supporting the
analysis of Connell and Temkin who \hfill have \hfill found their
\hfill model to have a wider

\begin{figure}
\vspace*{-0.5cm}
\hspace*{-1.2cm}
\epsfxsize=10cm
\epsfbox{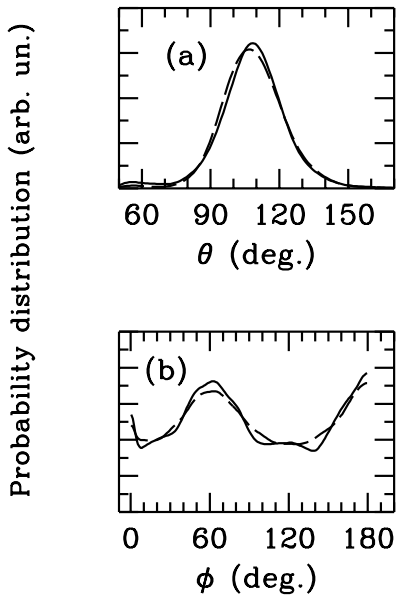}
\vspace*{-1.0cm}
\caption{
Distributions of (a) bond and (b) dihedral angles for CRN-A (dashed lines) and
CRN-B (full lines).
}
\label{fig:dihed}
\end{figure}

\noindent
distribution of bond angles than Polk, owing to the additional constraint on
the parity of atomic rings. However, we find here that CRN-A and CRN-B both
have a bond-angle-distribution of width about 11 degrees (at 0 K; cf.\ Table
\ref{tab:coord}), very much in agreement with experiment. Both distributions,
moreover, are centered closely on the ideal tetrahedral angle and are almost
Gaussian. The similarity between the angular distributions for two such
different models reflects the ability of the network to reorganize in spite
of topological constraints on the formation of odd-membered
rings.\cite{force}

We now turn to dihedral angles, i.e., angles between second-nearest-neighbor
bonds. Connell and Temkin found, in their model, staggered configurations
($\phi=60$ degrees) to be four times more numerous than in the Polk model,
concluding that this preference for staggered configurations should be a
signature of the absence of odd-membered rings, i.e., characteristic of
CRN-B-type materials. We find no support for such a conclusion here: as can
be seen in Fig.\ \ref{fig:dihed}(b), our two models exhibit essentially
identical dihedral-angle distributions.

As demonstrated in Ref.\ \onlinecite{mousseau97}, {\it a}-GaAs and {\it a}-Si
form networks which are topologically different. Structural signatures of
these differences, however, as we have seen here by comparing models CRN-A
and CRN-B, are extremely difficult to extract from experiment or even from
computer models; it seems to show up clearly, in fact, only in quantities
which cannot be measured directly, namely the number of wrong bonds or ring
statistics. Thus, most measurable quantities appear to be unaffected by the
constraint on wrong bonds. It should be noted that even though the
experimental precision required to decide between the two models 
on the basis of their total

\begin{figure}
\vspace*{-0.5cm}
\hspace*{-1.25cm}
\epsfxsize=10cm
\epsfbox{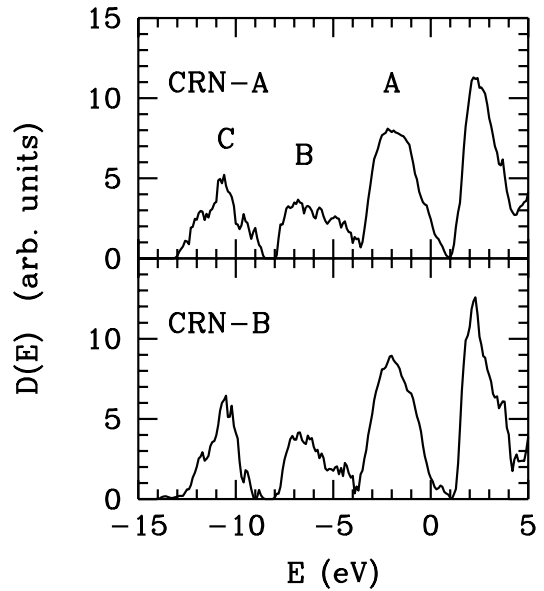}
\vspace*{-1cm}
\caption{
Electronic densities of states for the two models, as indicated. The
identification of the peaks is discussed in the text.
}
\label{fig:dos}
\end{figure}

\noindent
SSF's (or RDF's) can easily be achieved (cf.\ Figs.\
\ref{fig:fss} and \ref{fig:fdr}), the interpretation of the small differences
is not simple, as they could be due to variations in the modes of
preparation, details of the electronic potentials which could affect the
structure without changing the connectivity, etc. The problem could however be
resolved through measurements of the {\em partial} SSF's, as discussed above;
unfortunately, this does not seem to be possible at present.

\section{Electronic properties}

The electronic densities of states (DOS) for our two samples, CRN-A and
CRN-B, are displayed in Fig.\ \ref{fig:dos}. These were obtained by averaging
over the MD trajectories for a period of 3.5 ps at 300 K. As discussed in
Refs.\ \onlinecite{joannopoulos74} and \onlinecite{oreilly86}, the bands in
the DOS can be roughly ascribed as follows: the lowest-lying band, labeled
`C' in Fig.\ \ref{fig:dos}, is As $s$-like while the next one, `B', arises
from Ga $s$ and some As $p$ states; the gap between this band and the
following one is the ``ionicity'' gap. Just below the forbidden gap, band `A'
is composed of As $p$ and Ga $p$ states. In a crystal, the gap is direct and
has a width of 1.55 eV.

XPS measurements of amorphous GaAs have been interpreted as indicating that
the material is essentially chemically ordered.\cite{senemaud85} Probing the
valence band from the gap down to about $-$15 eV, XPS reveals relatively little
difference from the crystalline state except for the filling of the minimum
between the first (A) and second (B) peak and a shift of 0.5 eV upwards of
the third peak (C). However, upon comparing the DOS for our two networks
(Fig.\ \ref{fig:dos}), one sees an increased contribution in model CRN-A of
the high-energy side of the low-lying As $s$ band (C), almost forming an
additional peak; this is due to wrong As-As bonds, as was also shown in
Refs.\ \onlinecite{joannopoulos74} and \onlinecite{oreilly86}. The width of
the gap at 300 K between this band and the following mixed band, $-$1.5 eV in
the crystal, is reduced to about 0.9 eV in CRN-B and 0.5 eV in CRN-A.

Another manifestation of the presence of wrong bonds is visible in the
high-energy tail of the B band, which is much broader than in the
crystal.\cite{joannopoulos74,oreilly86} The added contributions at about
-4.5 eV {\em must} be due to wrong bonds since the structural properties of
the two models are very similar except for the number of such defects. This
is in agreement with the discussion presented in Ref.\ \onlinecite{seong96}.

It is clear from the comparison between the two models that although present,
the effects of the existence of wrong bonds on the electronic structure of
the material are much weaker than can hopefully be measured using techniques
such as XPS. Thus, the observed similarities in the XPS spectra of {\it a}-
and {\em c-}GaAs {\em cannot} be taken as evidence that the amorphous
material is chemically ordered,\cite{senemaud85} though our calculations do
show that this is indeed the case.

Disorder influences strongly the gap between valence and conduction bands.
The value for this quantity has been reported in the literature to lie in the
range 0.61--1.45 eV.\cite{semiconductor84} Although the gap in the CRN-A
sample is about 50\% smaller than the one for CRN-B, both are substantially
smaller than in the crystal. In particular, this value should depend
significantly on the method of preparation of amorphous GaAs and is probably
not a very good measure of the bond concentration.\cite{theye85}

\section{Vibrational properties}

\begin{figure}
\vspace*{-0.5cm}
\hspace*{-1.4cm}
\epsfxsize=10cm
\epsfbox{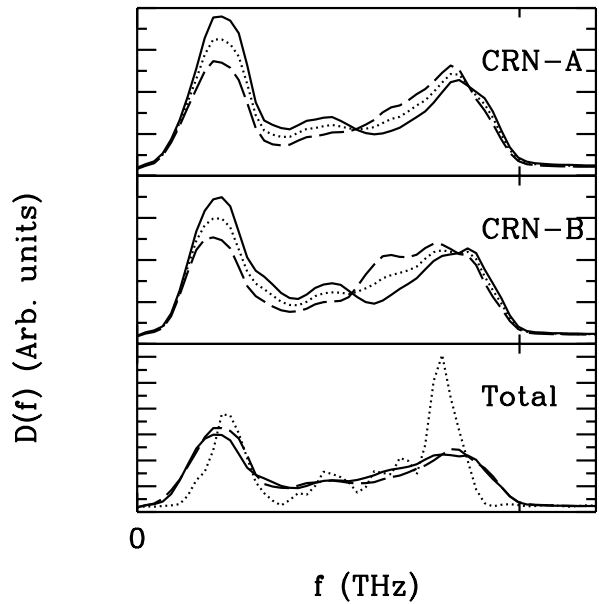}
\vspace*{-1.2cm}
\caption{
Partial and total vibrational densities of states for the two models, as
indicated. Dashed lines are for Ga atoms, solid lines for As and dotted lines
are the totals. The lower panel presents a comparison of CRN-A (dashed line)
and CRN-B (solid line) with the crystal (dotted line).
}
\label{fig:vdos}
\end{figure}

We have calculated the vibrational densities of states (VDOS) of our two
models by Fourier transforming the velocity auto-correlation function
averaged over 3.5 ps at 300 K; they are displayed in Fig.\ \ref{fig:vdos}.
Experimentally, this quantity can be extracted from Raman spectroscopy
measurements since amorphous networks do not have forbidden symmetries so
that all vibrational modes are allowed and measured. A multiple-order Raman
scattering study of {\it a}-GaAs has recently been
reported.\cite{chehaidar94} The VDOS extracted from these data reveals two
broad peaks, at about 2.3 and 8.3 THz, corresponding to the transverse
acoustic and optic modes, respectively, with a wide, almost featureless band
in between. At variance with the simulation results of Molteni {\it et
al.},\cite{molteni94b} the TO peak is very much present experimentally, with
a weight larger than that of the TA peak.

From our results --- Fig.\ \ref{fig:vdos} --- we see that the VDOS for our
two model networks are very similar: one difference is perhaps a slight shift
of weight from the TA to the TO

\noindent
peak for CRN-B compared to CRN-A. The great
similarity in the VDOS of our two models indicates that this quantity, also,
is rather insensitive to the presence of wrong bonds; this would not be the
case, however, if the two species differed appreciably in mass or elastic
properties. In a recent Car-Parrinello simulation of {\em a}-InP, for
instance, a high-energy peak in the partial P-P VDOS has been identified as
arising from wrong bonds;\cite{lewis97} because In is significantly heavier
than P, the corresponding peak for In-In is lost in the ``normal'' continuum
of states.

Although the agreement between simulation and experiment is satifactory, a
discrepancy remains regarding the relative amplitude of the two peaks. The
weight of the TA peak is slightly larger than that of the TO peak in our
simulation while the opposite is true experimentally. This difference
indicates that ``real'' {\it a}-GaAs would have even less coordination
defects that our model. Indeed, the TA peak is associated with bond-bending
modes which are relatively insensitive to the local order as long as the
stress is small, while the TO peak, depends critically on the existence of
the tetrahedral symmetry around each atom;\cite{mousseau91,alben75} any
coordination defect causes decrease of this peak. This is in fact clearly
evident if we compare our results with those of Molteni {\it et
al.},\cite{molteni94b} whose structure contains a much higher density of
coordination defects than ours and shows almost no TO peak. Based on this,
therefore, and on the agreement of our calculated DOS with experiment, we
conclude that real {\it a}-GaAs must be almost perfectly four-fold
coordinated.

\section{Concluding remarks}

Using a newly-proposed, event-based, Monte-Carlo optimization method, the
activation-relaxation technique of Barkema and Mousseau,\cite{barkema96} we
have constructed a model of {\it a}-GaAs with a minimum of wrong bonds
corresponding to an ``infinite'' Connell-Temkin model (CRN-B). This model is
found to be energetically favorable over the traditional Polk-type continuous
random network (CRN-A). The CRN-B model of {\it a}-GaAs represents, to the
best of our knowledge, the best realization to date of this material, with an
almost perfect fourfold coordination, realistic bond-angle distribution, and
almost no wrong bonds.

In order to provide insight into the structure of {\it a}-GaAs, a detailed
study of structural, electronic, and vibrational properties of CRN-B has been
presented, including a comparison with CRN-A. These results are in agreement
with experiment and suggest that ``real'' {\it a}-GaAs forms a perfect CRN
network, tetravalent and only weakly strained, with a minimum of wrong bonds
(ideally none). Our analysis also shows, however, that wrong bonds are
extremely difficult to identify experimentally; in particular, indirect
measurements (such as XPS and Raman scattering) cannot provide such
information. Likewise, diffraction experiments that do not discriminate
between the two chemical species are not sufficiently accurate to yield even
approximate estimates of the proportion of wrong bonds. Our calculations
indicate that only direct measurements of partial radial distribution
functions can provide experimental values for the density of wrong bonds.
Such measurements, unfortunately, do not seem to be possible at present. The
results presented here thus provide a useful reference for further
experimental and theoretical work.

\section{Acknowledgments}

We are grateful to G.\ T.\ Barkema for useful discussions. This work was
supported by grants from the Natural Sciences and Engineering Research
Council (NSERC) of Canada and the ``Fond pour la formation de chercheurs et
l'aide {\`a} la recherche'' of the Province of Qu{\'e}bec. Part of the
calculations were carried out at the ``Centre d'applications du calcul
parall{\`e}le de l'Universit{\'e} de Sherbrooke'' (CACPUS). We are grateful to
the ``Services informatiques de l'Universit{\'e} de Montr{\'e}al'' for generous
allocations of computer resources.

\bibliographystyle{prsty}

\begin{thebibliography}{99}
\bibitem{force} Relaxation of models CRN-A and CRN-B with the GSP TB
potential for Si yields a width of 9.44 and 9.70 degrees, respectively,
indicating that the bond bending force is a more important factor for
determining this quantity.
\end{thebibliography}

\end{document}